\documentclass[a4paper,conference]{IEEEtran}
\IEEEoverridecommandlockouts
\usepackage{cite}
\usepackage{amsmath,amssymb,amsfonts}
\usepackage{algorithmic}
\usepackage{graphicx}
\usepackage{textcomp}
\usepackage{psfrag}
\usepackage[lined,boxed,commentsnumbered]{algorithm2e}
\SetAlCapSkip{0.4em}
\usepackage{xcolor}
\usepackage{tikz}
\usetikzlibrary{positioning}
\def\BibTeX{{\rm B\kern-.05em{\sc i\kern-.025em b}\kern-.08em
    T\kern-.1667em\lower.7ex\hbox{E}\kern-.125emX}}
    
    \usepackage[absolute]{textpos}
\newcommand{\copyrightstatement}{
    \begin{textblock}{15}(0.5,0.3)    % tweak here: {box width}(leftposition, rightposition)
         \noindent
         \centering
         \textblockcolour{white}
         \scriptsize
         \copyright 2020 IEEE. Personal use of this material is permitted. Permission from IEEE must be obtained for all other uses, in any current or future media, including reprinting/republishing this material for advertising or promotional purposes, creating new collective works, for resale or redistribution to servers or lists, or reuse of any copyrighted component of this work in other works
    \end{textblock}
}

\begin{document}

\title{Matched Quality Evaluation of Temporally Downsampled Videos with Non-Integer Factors%\\
}

\copyrightstatement

\author{\IEEEauthorblockN{Christian Herglotz, Geetha Ramasubbu, Andr\'e Kaup}
\IEEEauthorblockA{\textit{Multimedia Communications and Signal Processing} \\ \textit{Friedrich-Alexander University Erlangen-N\"urnberg (FAU)},
Erlangen, Germany \\
\{christian.herglotz, geetha.ramasubbu, andre.kaup\}@fau.de}
%\and
%\and
%\and
%\and
%\and
\vspace{-1cm}
}
\IEEEoverridecommandlockouts 
\IEEEpubid{\makebox[\columnwidth]{978-1-7281-5965-2/20/\$31.00 \copyright 2020 IEEE \hfill} \hspace{\columnsep}\makebox[\columnwidth]{ }}

\maketitle

\begin{abstract}
Recent research has shown that temporal downsampling of high-frame-rate sequences can be exploited to improve the rate-distortion performance in video coding. %For example, depending on the content of the sequence, a higher rate-quality trade-off can be achieved when the frame rate is halved. 
%These results were verified using objective as well as subjective quality evaluations. 
However, until now, research only targeted downsampling factors of powers of two, which greatly restricts the potential applicability of temporal downsampling. A major reason is that traditional, objective quality metrics such as peak signal-to-noise ratio or more recent approaches, which try to mimic subjective quality, can only be evaluated between two sequences whose frame rate ratio is an integer value. To relieve this problem, we propose a quality evaluation method that allows calculating the distortion between two sequences whose frame rate ratio is fractional. The proposed method can be applied to any full-reference quality metric.  
\end{abstract}

\vspace{-0.5cm}
\begin{tikzpicture}[overlay, remember picture]
\path (current page.north) node (anchor) {};
\node [below=of anchor] {%
2020 Twelfth International Conference on Quality of Multimedia Experience (QoMEX)};
\end{tikzpicture}
\section{Introduction}
The world-wide demand for video streaming applications has led to the fact that nowadays, video data represents more than two thirds of the world-wide internet traffic \cite{cisco17}. To stream such a high amount of data, efficient video compression methods are needed to reduce the bitrates and, hence, requirements on transmission channels and their energy consumption. 

To achieve low bitrates, various video codecs such as High-Efficiency Video Coding (HEVC) \cite{ITU_HEVC}, H.264/AVC \cite{ITU_H.264}, or the upcoming Versatile Video Coding (VVC) were developed \cite{JVET_Q2001}. These codecs perform lossy compression % rely on redundancy and irrelevancy reduction. Redundancy reduction exploits prediction to avoid the necessity to transmit the same data multiple times. In contrast, irrelevancy reduction uses 
where quantization of coefficients is used to reduce the bitrate. Unfortunately, at the same time, the visual quality of the video suffers. Common to all these codecs is that during compression, the spatial and temporal resolution remains unchanged. %All these codecs have in common that during compression, the spatial as well as the temporal resolution of the video is not changed. 

Consequently, in the literature, it was tested whether downsampling of video data can be performed on top of quantization in order to increase the rate-distortion performance. In this direction, it was found that at a certain quantization step size, it is more beneficial to reduce the spatial resolution instead of further increasing quantization \cite{Wang14, Afonso17, Fischer20}. 
This approach could also be exploited to reduce the decoding energy \cite{Herglotz19b}. 

A similar observation was made for temporal downsampling, where, depending on the content of a sequence, reducing the frame rate is also beneficial for high quantization parameters \cite{Mackin17,Huang16, Ou11}. Subjective tests conducted with multiple viewers confirmed these findings. However, all these publications only consider temporal downsampling factors of 
\begin{equation}
d=\frac{f_\mathrm{ref}}{f_\mathrm{down}}=2^n, n\in \mathbb{N}, 
\end{equation} 
where $f_\mathrm{ref}$  is the frame rate of the original reference sequence, and $f_\mathrm{down}$ is the frame rate of the temporally downsampled sequence. 
Further examples of quality evaluation using integer downsampling are discussed in \cite{Ou11, Cheon16, Mackin17, Zhang17}. For example, the metric proposed by Ou et al. \cite{Ou11} assumes an exponential relation between the frame rate and the perceptual quality. However, it also relies on the traditional PSNR, and results were only evaluated on sequences downsampled with integer factors. Hence, a generalized approach to allow the calculation of PSNR and other quality metrics is needed.

To allow quality evaluation at non-integer downsampling factors, this paper presents an algorithm that calculates the distortion of videos downsampled with a factor of 
\begin{equation}
d=\frac{f_\mathrm{ref}}{f_\mathrm{down}}\in \mathbb{Q}, 
\label{eq:d_rational}
\end{equation}
which allows downsampling with any rational, possibly non-integer factor. As an example, a given sequence with a frame rate of $f_\mathrm{ref}=60$\,Hz can be compared to a temporally downsampled version with a frame rate of $f_\mathrm{down}=50$\,Hz. The idea is that the calculation of the quality metric (QM) is performed at the frame rate that corresponds to the least common multiple (LCM) of the two frame rates, which is $f_\mathrm{LCM}=300$\,Hz in the given example. %Hence, the proposed metric allows to perform objective quality evaluation with an extended range of downsampling factors. 

An important advantage of the proposed algorithm is that it can be applied to any traditional QM that performs full-reference distortion calculation, which means that the distortion is calculated by comparing the distorted frame with an undistorted reference frame. For example, the proposed algorithm can be applied to the classic peak signal-to-noise ratio (PSNR), but also to QMs mimicking subjective quality evaluation like the structural similarity index (SSIM) \cite{Wang04} or video multi-method assessment fusion (VMAF) \cite{VMAF}. To the best of our knowledge, these QMs could not be applied for non-integer downsampling factors as defined in \eqref{eq:d_rational} before. 

This paper is organized as follows. First, Section~\ref{sec:soat} explains how classical QMs were calculated using integer downsampling factors. Then, Section~\ref{sec:matched} introduces the proposed algorithm allowing QM calculation for non-integer downsampling. Afterwards, Section~\ref{sec:eval} presents some exemplary results. Section~\ref{sec:concl} concludes this paper.

\section{Quality Evaluation in Integer Downsampling}
\label{sec:soat}
\begin{figure}
  \centering
\psfrag{H}[r][r]{$f_\mathrm{ref}$}
\psfrag{L}[r][r]{$f_\mathrm{down}$}
\psfrag{t}[r][r]{Time $t$\,[s]}
	\includegraphics[width=.45\textwidth]{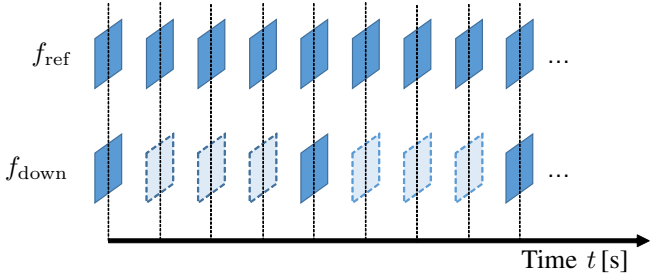}
		\vspace{-0.5cm}
	\caption{Padding technique for quality evaluation between two sequences with different frame rates. The frame rate of the reference sequence on the top is four times the frame rate of the downsampled sequence on the bottom ($f_\mathrm{ref}=4\cdot f_\mathrm{down}$).  }
\label{fig:intDownsampling}
\end{figure}

Traditionally, QMs for temporally downsampled sequences were calculated using a simple padding technique. This technique is illustrated in Fig.~\ref{fig:intDownsampling}. %, it is possible to calculate traditional full-reference quality metrics such as the PSNR or the SSIM. 

The figure shows two sequences (top and bottom) over the time on the horizontal axis. In this example, the frame rate of the top sequence is four times the frame rate of the bottom sequence. As such, four frames of the top sequence temporally correspond to one frame in the bottom sequence (illustrated by dark blue parallelograms). To enable quality evaluation, a single frame of the bottom sequence is repeated three times (light blue parallelograms), such that each frame of the high-frame-rate  reference sequence can be compared one by one to a corresponding frame of the low-frame-rate, downsampled sequence. Then, the QM is calculated for each frame of the reference sequence and averaged.

\section{Matched Quality Evaluation}
\label{sec:matched}
\begin{figure}
  \centering
  \providecommand\matlabtextB{\color[rgb]{0.950,0.2,0.2}}
\psfrag{H}[r][r]{$f_\mathrm{ref}$}
\psfrag{L}[r][r]{$f_\mathrm{down}$}
\psfrag{t}[r][r]{Time $t\,$[s]}
\psfrag{0}[c][c]{$0$}
\psfrag{8}[c][c]{$0.5$}
\psfrag{9}[c][c]{$1$}
\psfrag{1}[c][c]{\rotatebox{90}{\matlabtextB$1$}}
\psfrag{2}[c][c]{\rotatebox{90}{\matlabtextB$2$}}
\psfrag{3}[c][c]{\rotatebox{90}{\matlabtextB$3$}}
\psfrag{4}[c][c]{\rotatebox{90}{\matlabtextB$4$}}
\psfrag{5}[c][c]{\rotatebox{90}{\matlabtextB$5$}}
	\includegraphics[width=.45\textwidth]{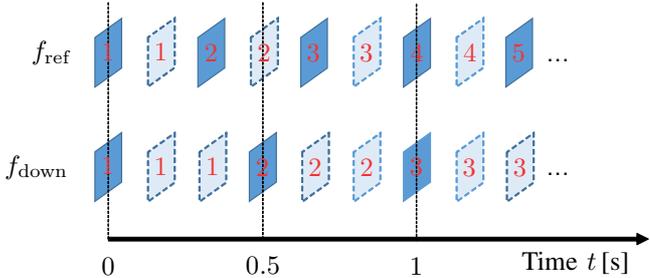}
	\vspace{-0.5cm}
	\caption{Time-stamps of real frames (dark blue) and virtual frames (light blue). The original frame indices are indicated by the red numbers.  }
\label{fig:nonIntSeqs}
\vspace{-0.4cm}
\end{figure}

To enable quality evaluation between two sequences with a frame rate ratio of $d=\frac{f_\mathrm{ref}}{f_\mathrm{down}}\in \mathbb{Q}$, we introduce the concept of a virtual frame. Similar to the padded frames in Fig.~\ref{fig:intDownsampling}, virtual frames are also padded frames. In contrast to integer downsampled quality evaluation, padding is performed in both sequences. Figure~\ref{fig:nonIntSeqs} shows an example with a reference sequence of $3\,$Hz and a downsampled sequence of $2\,$Hz. 

We can see that only each second, the time stamps of the reference as well as the downsampled frames (dark blue) match. In between, we introduce virtual frames (light blue) to enable matching time stamps. In our proposed method, these virtual frames are copies from the preceding frame (padding), which is indicated by the frame index in red. The resulting frame rate $f_\mathrm{LCM}$ for both sequences is the LCM of the two original frame rates $f_\mathrm{ref}$ and $f_\mathrm{down}$. This padding technique does not change the content of both sequences because a viewer would not be able to distinguish the sequence at its original frame rate  from the padded sequence at the frame rate $f_\mathrm{LCM}$. The reason is that a screen displays each frame until the next frame's time stamp, such that copied frames at intermediate time stamps do not change the visual content. % occurs. %time stamp of the next frame. 

We further introduce the concept of a cluster, which is defined as the time range between two temporally matching frames at the original frame rates. In the example, the cluster has a duration of $1\,$s with frame numbers per cluster %, which corresponds to frame numbers per cluster of 
\begin{equation}
N_\mathrm{ref} = \frac{f_\mathrm{ref}}{\mathrm{GCD}(f_\mathrm{ref},f_\mathrm{down})},\quad N_\mathrm{down} = \frac{f_\mathrm{down}}{\mathrm{GCD}(f_\mathrm{ref},f_\mathrm{down})}
\end{equation} 
for the reference and the downsampled sequence, respectively. The operator GCD represents the greatest common divisor of both frame rates. In the example, the cluster lengths yield $N_\mathrm{ref}=3$ and $N_\mathrm{down}=2$. For a given duration $T$\,[s] of the two sequences, we can calculate the number of clusters 
\begin{equation}
C = \frac{T\cdot f_\mathrm{ref}}{N_\mathrm{ref}}= \frac{T\cdot f_\mathrm{down}}{N_\mathrm{down}}. 
\end{equation}

From the two cluster lengths $N_\mathrm{ref}$ and $N_\mathrm{down}$, we can calculate the virtual cluster length
\begin{equation}
N_\mathrm{V} = N_\mathrm{ref}\cdot N_\mathrm{down}, 
\end{equation}
which corresponds to the sum of real frames and the virtual frames within a cluster. %, which is valid for both sequences. %This number is the same for both high-frame-rate and low-frame-rate sequences. 
In the virtual domain, the classical distortion measures can now be calculated for non-matching time stamps by comparing real frames to virtual frames.  

To minimize QM calculation times, we take into account that the overlap of frames in the time domain is not constant. In the given example, the first frames of both reference and downsampled sequences overlap at $t=0\,$s and $t=\frac{1}{6}\,$s, which corresponds to two time stamps at a frame rate of $f_\mathrm{LCM}=6\,$Hz. In contrast, the next frame of the reference sequence only overlaps once with the first frame of the downsampled sequence. In our implementation, we consider this effect by calculating the QM only once between every co-occurring frame, weighted with the number of co-occurring time stamps. 

The weight vector $\boldsymbol{w}$ as well as the corresponding frame index vectors $\boldsymbol{h}$ and $\boldsymbol{l}$ for a single cluster are obtained using Algorithm~\ref{alg:weightGen}. 
\begin{algorithm}[t]
\label{alg:weightGen}
\SetKwInOut{Input}{input}\SetKwInOut{Output}{output}
\Input {$f_\mathrm{ref},f_\mathrm{down},N_\mathrm{ref}, N_\mathrm{down}$}
\Output {$\boldsymbol{w}, \boldsymbol{h}, \boldsymbol{l}$}
lastStamp$\,\gets 0$; lastFrame$_\mathrm{ref}$, lastFrame$_\mathrm{down}\gets$ 1\;% \%Init support variables\\
$\boldsymbol{w}, \boldsymbol{l}, \boldsymbol{h}\gets \boldsymbol{0}$\;% \%Init vectors \\
%\% Loop over all neighbors
\For {$i \gets \{1,2,...,(N_\mathrm{ref}+N_\mathrm{down}-1)\}$}{
$\boldsymbol{l}(i)\gets $lastFrame$_\mathrm{down}$\; %\%Set frame indices of $i$-th QM \\
$\boldsymbol{h}(i)\gets $lastFrame$_\mathrm{ref}$\; %\%calculation \\
%\%Test which sequence causes the next frame boundary\\
\eIf(){$N_\mathrm{down}\cdot\mathrm{lastFrame}_\mathrm{ref}<N_\mathrm{ref}\cdot\mathrm{lastFrame}_\mathrm{down}$}
{ $\mathrm{nextStamp}\gets N_\mathrm{down}\cdot\mathrm{lastFrame}_\mathrm{ref}$\;
	$\mathrm{lastFrame}_\mathrm{ref}\gets \mathrm{lastFrame}_\mathrm{ref}+1$\;}
	{$\mathrm{nextStamp}\gets N_\mathrm{ref}\cdot\mathrm{lastFrame}_\mathrm{down}$\;
	$\mathrm{lastL}\gets \mathrm{lastFrame}_\mathrm{down}+1$\;}
	$\boldsymbol{w}(i)\gets \mathrm{nextStamp}-\mathrm{lastStamp}$\;
	$\mathrm{lastStamp}\gets\mathrm{nextStamp}$\;
}
\caption{\small{Generation of weights and indices for QM calculation.  The number of iterations in the loop $N_\mathrm{ref}+N_\mathrm{down}-1$ is the number of required QM calculations per cluster.}}
\end{algorithm}
To give an example, the resulting weight and index vectors for the example of $f_\mathrm{ref}=3\,$Hz and $f_\mathrm{down}=2\,$Hz are $\boldsymbol{w} = \{2,1,1,2\}$, $\boldsymbol{h} = \{1,2,2,3\}$, and $\boldsymbol{l} = \{1,1,2,2\}$. The weights in $\boldsymbol{w}$ add up to the virtual cluster length $N_\mathrm{V}$.  % and that a minimum number of QM calculations must be performed. %listed in Table~\ref{tab:exampleVectors}. 
%\begin{table}[t]
%\renewcommand{\arraystretch}{1.3}
%\caption{Content of the vectors $\boldsymbol{w}, \boldsymbol{l}, \boldsymbol{h}$ after execution of Algorithm~\ref{alg:weightGen} with $f_\mathrm{ref}=3\,$Hz and $f_\mathrm{ref}=2\,$Hz. }
%\label{tab:cpPowers}
%\begin{center}
%\begin{tabular}{r||c|c|c|c}
%$\boldsymbol{w}$ & $2$ & 1 & 1 & 2 \\
%$\boldsymbol{l}$\\
%$\boldsymbol{h}$\\
% \end{tabular}
%\end{center}
%\end{table}

Finally, the overall QM of the complete sequence $q$ is calculated using Algorithm~\ref{alg:QMcalc}, where $q_\mathrm{C}$ is the QM for a single cluster. 
\begin{algorithm}[t]
\label{alg:QMcalc}
\SetKwInOut{Input}{input}\SetKwInOut{Output}{output}
\Input {$N_\mathrm{ref}, N_\mathrm{down}, \boldsymbol{w}, \boldsymbol{h}, \boldsymbol{l}, C$}
\Output {QM}
$o_\mathrm{ref}, o_\mathrm{down},q \gets 0$\;
\For {$n=\{1,2,...,C\}$}
{
	$q_\mathrm{C}\gets 0$\;
	\For{$i=\{1,2,...,(N_\mathrm{ref}+N_\mathrm{down}-1)\}$}
	{
		$q_\mathrm{C} \gets q_\mathrm{C} + \boldsymbol{w}(i)\cdot \mathrm{QM}\big(\boldsymbol{h}(i)+o_\mathrm{ref},\boldsymbol{l}(i)+o_\mathrm{down}\big)$\;
	}
	$o_\mathrm{ref}\gets o_\mathrm{ref} + N_\mathrm{ref}$\;
	$o_\mathrm{down}\gets o_\mathrm{down} + N_\mathrm{down}$\;
	$q\gets q+\frac{q_\mathrm{C}}{N_\mathrm{V}}$\;
}
$q\gets\frac{q}{C}$\;
\caption{\small{QM calculation over all clusters. The variables $o_\mathrm{ref}$ and $o_\mathrm{down}$ are offset indices to the current cluster in the reference and downsampled sequence, respectively. QM(.) is the quality metric operator, i.e., the frame-wise distortion between the frames in the left index and the right index.  }\vspace{-0.5cm}}
\end{algorithm}
The algorithm loops over all clusters and retrieves the original frame indices using the offset values $o_\mathrm{ref}$ and $o_\mathrm{down}$ plus the indices from the cluster-domain frame index vectors $\boldsymbol{h}$ and $\boldsymbol{l}$. The final QM $q$ is the time-average over all clusters. In the case of integer downsampling factors or no temporal downsampling, the proposed algorithm automatically reverts to classic QM calculations.

\section{Evaluation}
\label{sec:eval}
To show results, we apply the proposed algorithm to the QMs PSNR, SSIM, and  VMAF. Due to space constraints, we evaluate one sequence from the BVI-HFR dataset proposed in \cite{Mackin15} that provides multiple sequences at a frame rate of $120\,$Hz. %Downsampling was performed using frame dropping, where 
Target frame rates were set to $f_\mathrm{down}\in\{100, 60, 50, 40, 30, 25, 24, 15\}$, which corresponds to the downsampling factors $d$ listed in Table~\ref{tab:dScores}. %$\in\{\frac{6}{5}, 2, \frac{12}{5},3,4,\frac{24}{5},5,8\}$. 
Temporal downsampling was performed using frame averaging. %dropping, where the frame with the same time stamp or otherwise the temporally closest preceding frame was kept. 

%Figure~\ref{fig:exampleDown} 
Table~\ref{tab:dScores} shows results for the proposed matched PSNR (mPSNR), mSSIM, and mVMAF applied on a sequence with simple motion (\textit{golf\_side} from \cite{Mackin15}). 
\begin{table}[t]
\renewcommand{\arraystretch}{1.3}
\caption{mPSNR, mSSIM, and mVMAF score for different frame rates of the sequence \textit{golf\_side}. }
\vspace{-0.5cm}
\label{tab:dScores}
\begin{center}
\begin{tabular}{r|c||r|r|r}
\hline
$f_\mathrm{down}$\,[Hz] & $d$ & mPSNR [dB] & mSSIM & mVMAF\\
\hline
$120$ &  $1$   &  $\infty$    &  $1.000$   &  $97.43$     \\
$100$ &  $\frac{6}{5}$   &  $44.89$    &  $0.9981$   &   $94.56$    \\
$60$ &   $2$   &  $44.30$   &   $0.9976$  &    $93.51$   \\
$50$ &   $\frac{12}{5}$   &  $40.72$   &   $0.9960$  &     $92.92$  \\
$40$ &   $3$   &  $40.67$   &   $0.9955$  &    $92.62$   \\
$30$ &   $4$   &  $39.14$   &   $0.9936$  &    $91.48$   \\
$25$ &   $\frac{24}{5}$   &  $36.57$   &   $0.9916$  &    $90.75$   \\
$24$ &   $5$   &  $37.49$   &   $0.9918$  &    $90.56$   \\
$15$ &   $8$   &  $33.91$   &   $0.9865$  &    $88.25$   \\
\hline
 \end{tabular}
\end{center}
 %\vspace{-0.6cm}
\end{table}
The numbers show that as expected, all QM values drop when the frame rate is decreased. %Interestingly, a frame rate of $24\,$Hz, which is an integer downsampling factor of $d=5$, returns slightly higher QM values than a frame rate of $25\,$Hz, which corresponds to non-integer downsampling. 

\begin{figure}
\vspace{-0.1cm}

  \centering
\providecommand\matlabtextA{\color[rgb]{0.150,0.150,0.150}}%
\psfrag{013}[tc][tc]{\matlabtextA File size [Mb]}%
\psfrag{014}[bb][bc]{\matlabtextA mVMAF score}%

\providecommand\matlabtextB{\color[rgb]{0.150,0.150,0.150}}%
\psfrag{000}[tc][tc]{}%\matlabtextB $\times10^{6}$}%

\psfrag{001}[ct][ct]{\matlabtextB $0$}%
\psfrag{002}[ct][ct]{\matlabtextB $2$}%
\psfrag{003}[ct][ct]{\matlabtextB $4$}%
\psfrag{004}[ct][ct]{\matlabtextB $6$}%
\psfrag{005}[ct][ct]{\matlabtextB $8$}%
\psfrag{006}[ct][ct]{\matlabtextB $10$}%
\psfrag{007}[ct][ct]{\matlabtextB $12$}%

\psfrag{008}[rc][rc]{\matlabtextB $75$}%
\psfrag{009}[rc][rc]{\matlabtextB $80$}%
\psfrag{010}[rc][rc]{\matlabtextB $85$}%
\psfrag{011}[rc][rc]{\matlabtextB $90$}%
\psfrag{012}[rc][rc]{\matlabtextB $95$}%
\psfrag{data1aaaaa}[ll][ll]{ $120\,$Hz}%
\psfrag{data2}[ll][ll]{ $100\,$Hz}%
\psfrag{data3}[ll][ll]{ $60\,$Hz}%
\psfrag{data4}[ll][ll]{ $25\,$Hz}%
\psfrag{data5}[ll][ll]{ $24\,$Hz}%
	\includegraphics[width=.41\textwidth]{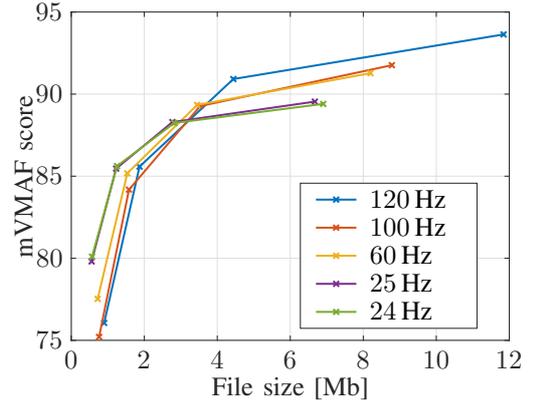}
	\vspace{-0.3cm}
	\caption{mVMAF score and bitrates for the sequence \textit{golf\_side} \cite{Mackin15} downsampled to the frame rates shown in the legend and compressed with different crf values. }
\label{fig:compressed}
%\vspace{-0.5cm}
\end{figure}

The matched QM method can also be applied to video compression. To this end, we present another example using HEVC-coded sequences. We encode the temporally downsampled sequence \textit{golf\_side} using x265 \cite{x265} with four constant rate factors (crf) $18,23,28$, and $33$. The resulting rate-distortion curves for the downsampled sequences evaluated with mVMAF are shown in Fig.~\ref{fig:compressed}. 
The curves confirm that for low bitrates, it can be beneficial to reduce the frame rates because the rate-distortion curves intersect. 

Finally, it is important to note that the observable quality scores highly depend on the content of the sequence. For example, for high-motion sequences, it can occur that rate-distortion curves do not intersect, which means that the original frame rate is always best in terms of rate-distortion performance. 

\vspace{-0.4cm}
\section{Conclusions}
\label{sec:concl}

This paper proposed a novel evaluation method for temporally downsampled sequences using traditional full-reference quality metrics. The method allows for calculating the distortion for fractional downsampling factors. The results of the evaluation indicate that for video compression, it is potentially beneficial to reduce the frame rate in order to increase the rate-distortion performance. 

Further work shall verify our results using subjective quality evaluation. Furthermore, the method shall be applied in video compression to optimize rate-distortion performance and reduce the decoding energy. 
%
%\section*{Acknowledgment}
%
%The preferred spelling of the word ``acknowledgment'' in America is without 
%an ``e'' after the ``g''. Avoid the stilted expression ``one of us (R. B. 
%G.) thanks $\ldots$''. Instead, try ``R. B. G. thanks$\ldots$''. Put sponsor 
%acknowledgments in the unnumbered footnote on the first page.

\bibliographystyle{IEEEtran}
\bibliography{D:/Literatur/literatureNeu}

\end{document}